\documentclass[aps,prl,amsmath,twocolumn,superscriptaddress,showpacs,nofootinbib]{revtex4}

\usepackage{fontenc}
\usepackage{amssymb}
\usepackage{amsmath}
\usepackage{amsfonts}
\usepackage{physics}
\usepackage{graphicx}
\usepackage{color}
\usepackage[normalem]{ulem}
\usepackage{bm}
\usepackage[percent]{overpic}
\let\vec\bm

\newcounter{myparagraphs}

\begin{document}
\title{Dissipative failure of adiabatic quantum transport as a dynamical phase transition}
\author{F. Barratt}
\affiliation{Department of Mathematics, King's College London, Strand, London WC2R 2LS, United Kingdom}

\author{Aleix Bou Comas}
\affiliation{Graduate Program in Physics and Initiative for the Theoretical Sciences, Graduate Center, CUNY, New York, NY 10016, USA}
\affiliation{Department of Physics and Astronomy, College of Staten Island, CUNY, Staten Island, NY 10314, USA}

\author{P. Crowley}
\affiliation{Department of Physics, Massachusetts Institute of Technology, Cambridge, Massachusetts 02139, USA}
%\affiliation{Department of Physics, Boston University, MA 02215}

\author{V. Oganesyan}
\affiliation{Graduate Program in Physics and Initiative for the Theoretical Sciences, Graduate Center, CUNY, New York, NY 10016, USA}
\affiliation{Department of Physics and Astronomy, College of Staten Island, CUNY, Staten Island, NY 10314, USA}

\author{P. Sollich}
\affiliation{Department of Mathematics, King's College London, Strand, London WC2R 2LS, United Kingdom}
\affiliation{Institute for Theoretical Physics, University of G\"ottingen, Friedrich-Hund-Platz 1, D-37077 G\"ottingen, Germany}

\author{A.~G. Green}
\affiliation{London Centre for Nanotechnology, University College London, Gordon St., London, WC1H 0AH, United Kingdom}

\date{\today}
\begin{abstract}
Entanglement is the central resource in adiabatic quantum transport. 
Dephasing affects the availability of that resource by biasing trajectories, driving transitions between success and failure. This depletion of entanglement is important for the practical implementation of quantum technologies.
We present a new perspective on the failure of adiabatic computation by understanding the failure of adiabatic transport as a dynamical phase transition. These ideas are demonstrated  in a toy model of adiabatic quantum transport in a two spin system.
\end{abstract}
\maketitle

Adiabatic transport is a powerful way to prepare quantum states. 
A system in the groundstate of a simple Hamiltonian can be transformed to a more complicated state by slowly and continuously changing its Hamiltonian to one for which the desired state is the groundstate~\cite{born1928adiabatic}. 
This approach is frequently used to prepare correlated states of cold atomic gases~\cite{bergmann1998k}. 
By encoding the result of a computation in the final state, it may also be used for quantum computation~\cite{kadowaki1998quantum,brooke1999j,farhi2001quantum,santoro2002theory}.

Adiabatic quantum computation (AQC) has been demonstrated to be computationally equivalent to gate-based quantum computation~\cite{Aharonov2008adiabatic} and various works have aimed at delimiting the classes of problem for which AQC succeeds or fails ~\cite{Altshuler2010,Jorg2010,laumann2015quantum}. 
A practically pressing question is how coupling to the environment causes a computation that would succeed in a pure system to fail.

Several approaches have been developed to consider these environmental effects on AQC. 
Viewed in the computational basis, one may study tunnelling between computational states~\cite{Denchev2016}. An alternative  --- adopted here --- is to determine the entanglement resources that can be maintained in the presence of the environment~\cite{Crowley2014,bauer2015entanglement}. 
Similar effects can also be captured by environmental renormalisation of the system's gap structure~\cite{wild2016adiabatic}.

Computation is a dynamical process. 
The transition between successful and unsuccessful computation is a transition in those dynamics, caused by a biasing of computational trajectories by environmental dephasing. 
In gate-based computation, there exist threshold strengths of dephasing that can be completely corrected for by suitable error correction~\cite{aharonov2008threshold}. 
Although error correction schemes have been proposed~\cite{jordan2006error,young2013error} and demonstrated~\cite{pudenz2014error} for AQC, no such thresholds are known and new perspectives are  evidently required.
We demonstrate that, for a simple model, the environment-induced failure of the adiabatic process can be understood as a dynamical phase transition using trajectory ensemble methods developed in the field of spin glasses.

{\bf A Simple Adiabatic Process:}
%%%%%%%%%%%%%%%%%%%%%%%%%%%%%%%%%
%
\begin{figure}[t]
    \centering
    \includegraphics[width=0.8\linewidth]{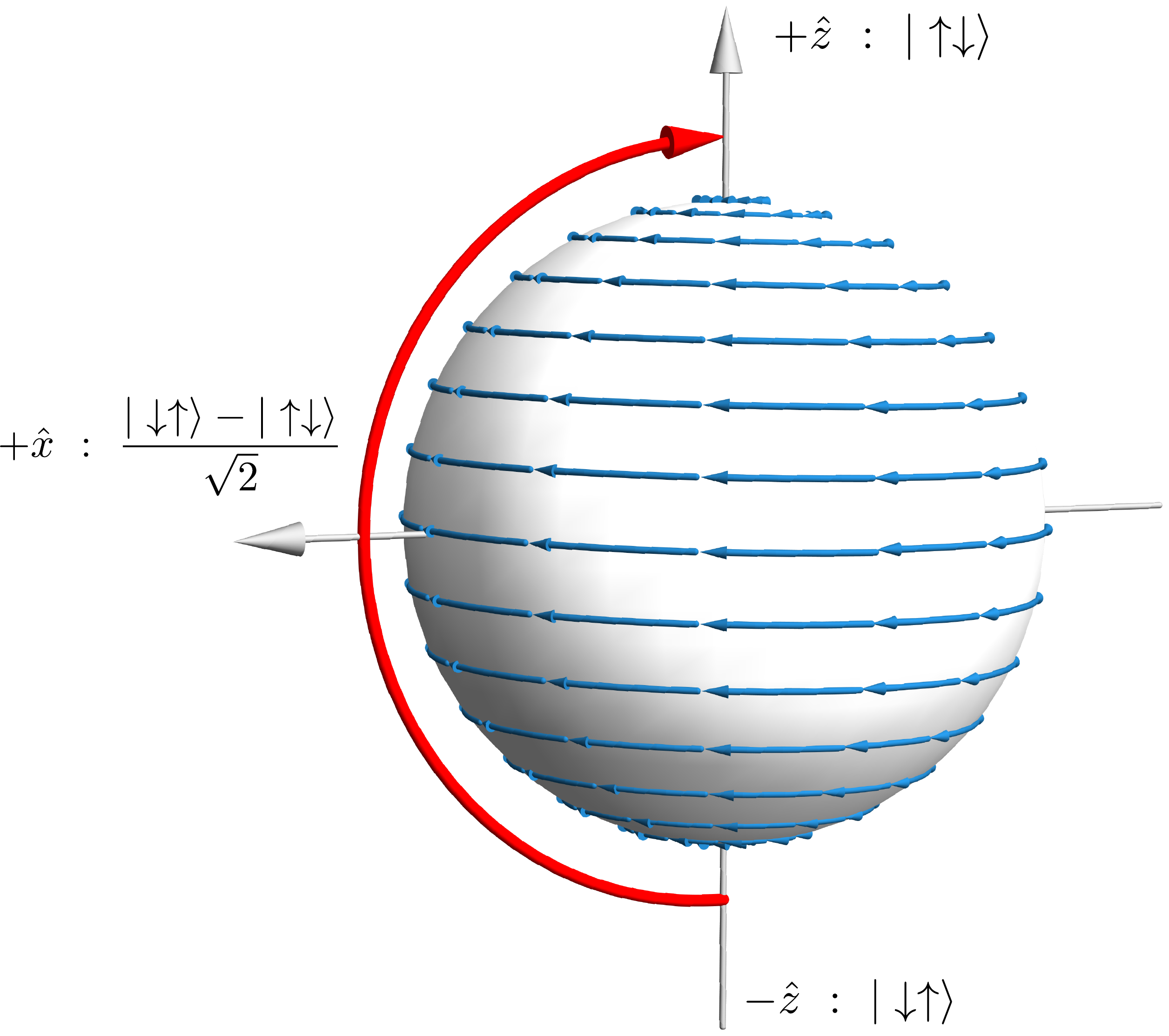}
    \caption{{\it A simple adiabatic quantum transport: } A
             subset of the entangled states of 2 spins form a Bloch sphere: $\cos(\theta/2)e^{i\varphi/2} \ket{\uparrow\downarrow} + \sin(\theta/2)e^{-i\varphi/2} \ket{\downarrow\uparrow}$ .  
             The ground state of $\hat{H} = \frac{J}{2} \hat{\vec{\sigma}}_1\cdot \hat{\vec{\sigma}}_2 + h(t) (\hat{\sigma}_1^z-\hat{\sigma}_2^z)$ as $h(t)$ is scanned from $-\infty$ to $\infty$ follows the path shown here (red arrow), passing through a maximally entangled state. 
             Flow fields of the $t=0$ hamiltonian are shown in blue.}\label{fig:adiabatic}
\end{figure}
The success of quantum adiabatic transport depends upon the dynamics of its entanglement resources.
The simplest model of this is two coupled quantum spins-$\frac{1}{2}$. 
We study a simple adiabatic process of evolution under an antiferromagnetic Heisenberg model with staggered, time-dependent field:
 \begin{equation}
     H = \frac{J}{2} \vec{\hat{\sigma}}_1 \cdot \vec{\hat{\sigma}}_2 + h(t)(\hat{\sigma}_1^z-\hat{\sigma}_2^z).
\label{FullH}
\end{equation}
The system is initialised in the state $|\uparrow \downarrow \rangle $ and the field swept from $h(0)=-\infty$ to $h(T)=\infty$.  
Our analysis is conveniently carried out in terms of the following parametrization of the two-spin Hilbert space:
\begin{equation}
\ket{\psi} = n_{1}\ket{\vec{l}_1, \vec{l}_2} + n_{2}\ket{-\vec{l}_1, -\vec{l}_2},
\label{parametrisation}
\end{equation}
where the $\ket{\vec{l}_i}$ are spin coherent states, and $\ket{-\vec{l}_i}$ is the state such that $\braket{\vec{l}_i}{-\vec{l}_i}=0$. 
The dynamics with this parametrization is particularly simple: the vectors $\vec{l}_1=-\vec{l}_2=\hat {\vec{z}} $ do not evolve and the state of the system becomes $\ket{\psi} = n_{1}\ket{\uparrow \downarrow} + n_{2}\ket{\downarrow \uparrow}\equiv \cos \theta/2\ket{\uparrow \downarrow} + e^{i \phi}\sin \theta/2 \ket{\downarrow \uparrow} $, where we have represented the entanglement spinor $(n_1,n_2)$ as a point on a Bloch-like sphere following standard convention.
In other words, the state evolves in a reduced two-dimensional subspace of the entire 4-dimensional Hilbert space -- Fig. \ref{fig:adiabatic} shows the adiabatic path on this Bloch sphere. 
In the zero-magnetization subspace, the Schr\"odinger equation
%%%%%%%%%%%%%%%%%%%%%%%%%%%%%%%%%%%%
reduces to the \emph{classical} equation of motion 
for the Bloch unit vector $
\dot{\vec{n}} = \left[J \hat{x} + h(t) \hat{z}\right] \cross \vec{n}.
$
We can also identify a vector of operators
\begin{equation}
    \hat{\vec{\tau}} = \begin{pmatrix} \hat{\tau_x} \\ \hat{\tau_y} \\ \hat{\tau_z} \end{pmatrix} = \begin{pmatrix}(\hat{\sigma}_1^+\hat{\sigma}_2^-+\hat{\sigma}_1^-\hat{\sigma}_2^+) \\ -i(\hat{\sigma}_1^+\hat{\sigma}_2^- - \hat{\sigma}_1^-\hat{\sigma}_2^+) \\ \flatfrac{(\hat{\sigma}_2^z-\hat{\sigma}_1^z)}{2} \end{pmatrix},
    \label{eq:tau}%used ot be eq:n
\end{equation}
that obey $su(2)$ commutation relations, in terms of which 
$\hat H = J\left(\hat \tau_x - \hat \tau_z^2 + \frac{1}{2}\right)-2h(t) \hat\tau_z$.
In our model of adiabatic transport, the ability to sustain entanglement at the instant when $h=0$ determines whether the trajectory is connected and so whether the process is successful. We therefore investigate dynamics directly at $h=0$, i.e. $\hat H = J \hat \tau_x$ (for pseudospin 1/2) henceforth.

{\bf Introducing Dissipation:}
%%%%%%%%%%%%%%%%%%%%%%%%%%%%%%%%%%%%%%%%%%%%%%
\begin{figure}[t]
    \includegraphics[width=\linewidth]{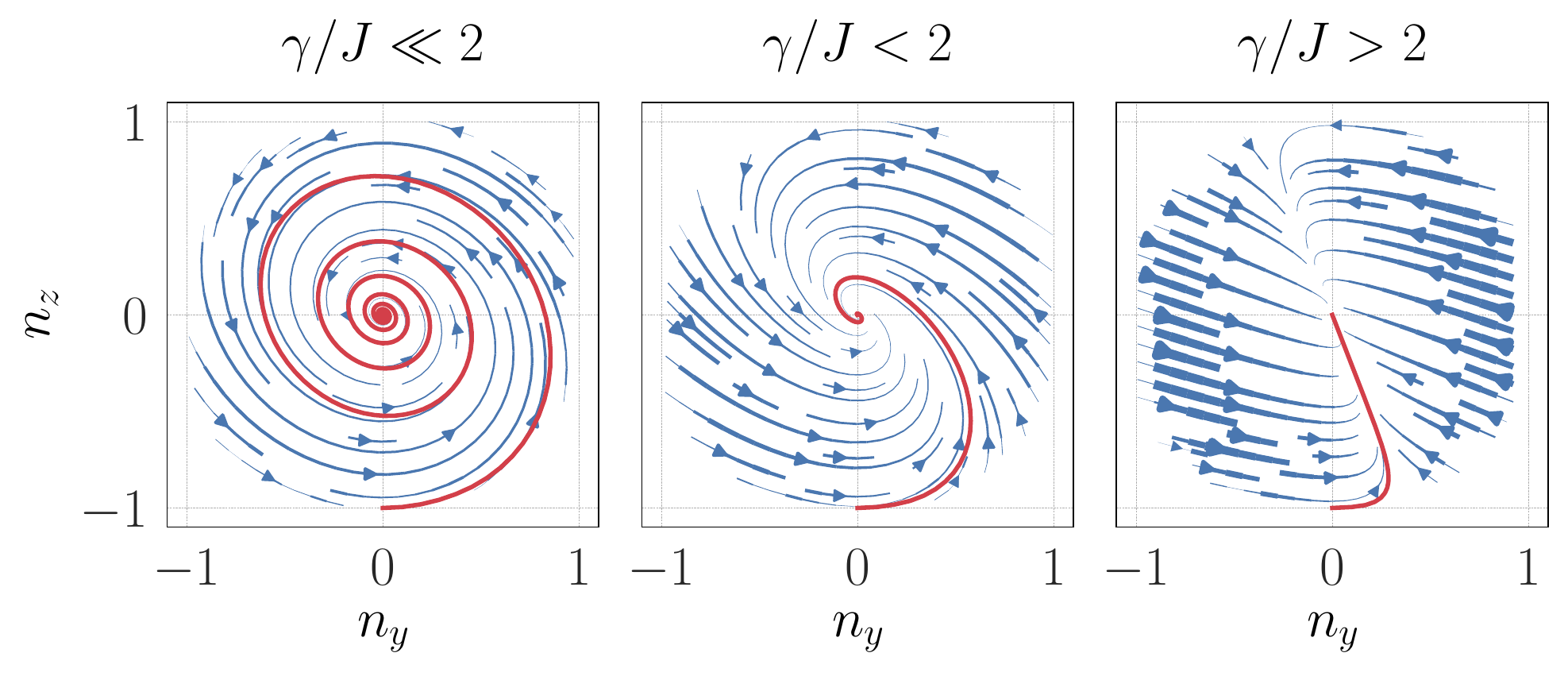}
    \caption{{\it Flow fields and trajectories in the presence of dissipation:} Blue lines show a cut through the flow fields for Eq.~(\ref{eq:lindblad}) along the $z$-$y$ plane for different dissipation strengths. Red curves indicate the  fate of trajectories starting at $\ket{\uparrow\downarrow}$. $\gamma/J<2$ trajectories explore both hemispheres; $\gamma/J>2$ trajectories remain in the bottom hemisphere.}\label{fig:av_diss}
    \label{fig:underover}
\end{figure}
We model the  environment using harmonic baths coupled locally to each spin. This assumption of locality --- and the corollary that the number of dissipation channels is proportional only to the number of spins --- is physically reasonable and underpins the possibility of performing quantum error correction. 
We consider random fields only along $\hat{z}$, motivated by systems in which different components of the qubit are of different physical origin, {\it e.g.} a flux qubit with noise arising from inductive coupling to circulating currents.
If treated in a Keldysh formalism, the environment can be modelled by a random noise and a corresponding friction, resulting in a modified Schr\"odinger equation~\cite{Crowley2016,kamenev2011field}:
\begin{eqnarray*}
    i\partial_t \ket{\psi(t)} &=& \Big[\hat{H} + \sum_{i=1, 2}\big(\eta_i^z(t)\hat{\sigma}_i^z 
    \\
    & &+ \int_0^t\dd{t'} \Gamma(t-t') \dot{\ev{\sigma_i^z}}_{t'} \hat{\sigma}_i^z\big)\Big]\ket{\psi(t)}.
\end{eqnarray*}
Correlations in the noise fields $\eta^z_i(t)$ are related to the dissipation kernel $\Gamma(t-t')$ by the fluctuation-dissipation relation. 
We study the Markovian limit $\Gamma(t-t') = \Gamma\delta(t-t')$. 
Starting from 
$\ket{\uparrow\downarrow}$, the $\vec{l}_i$ fields do not change even when coupled to the environment. The Schr\"odinger equation for the entanglement field reduces to
\begin{equation}
\dot{\vec{n}} = \left[J \hat{x} + (h(t)+\tilde{\eta}(t)) \hat{z} -2\Gamma \dot{\vec{n}} \right] \cross \vec{n},
\label{eq:noiseyEq}
\end{equation}
for $\vec{n}=\langle\hat{\vec{\tau}}\rangle$. The effective noise field, $\tilde{\eta} = \eta_1-\eta_2$ has twice the variance of the local noises. Here, all stochastic differential equations should be understood as Stratonovich SDEs.

Fundamentally, it is dephasing that limits the availability of quantum resources. Friction can be systematically corrected for by applying appropriate drives or other compensating control to counter its effects. Dephasing cannot be corrected for in this way. Therefore,
we will ignore the effect of friction.
This amounts to a high temperature limit; $\Gamma\rightarrow 0$, $T \rightarrow \infty$, $\Gamma T \rightarrow \gamma$ finite.

{\bf The Effect of Local Dissipation:}
%%%%%%%%%%%%%%%%%%%%%%%%%%%

\noindent
{\it Averaged Dynamics:}
%%%%%%%%%%%%%%%
The average over noise can be performed after converting Eq.~(\ref{eq:noiseyEq}) to an Ito SDE and allowing the state %Bloch 
vector (now denoted $\bar{\vec n}$) to explore the interior of the Bloch sphere resulting in 
\begin{align}
    \dot{\bar{\vec{n}}} &= J \hat{x} \cross \bar{\vec{n}} - \gamma \hat{z} \cross (\bar{\vec{n}} \cross \hat{z}).
\label{eq:lindblad}
\end{align}
This equation is equivalent to the Heisenberg picture Lindblad equation for the operators $\hat{\vec{\tau}}$ of Eq.~(\ref{eq:tau}). It has a single fixed point at the origin that is stable for all values of the coupling. This linear problem exhibits an ``underdamped'' to ``overdamped'' spectral transition in the dynamics near the fixed point at $\gamma/J=2$, as illustrated in Fig. \ref{fig:underover}. Overdamped dynamics is confined to the lower hemisphere.
\begin{figure}[t]
\centering
    \includegraphics[width=\linewidth]{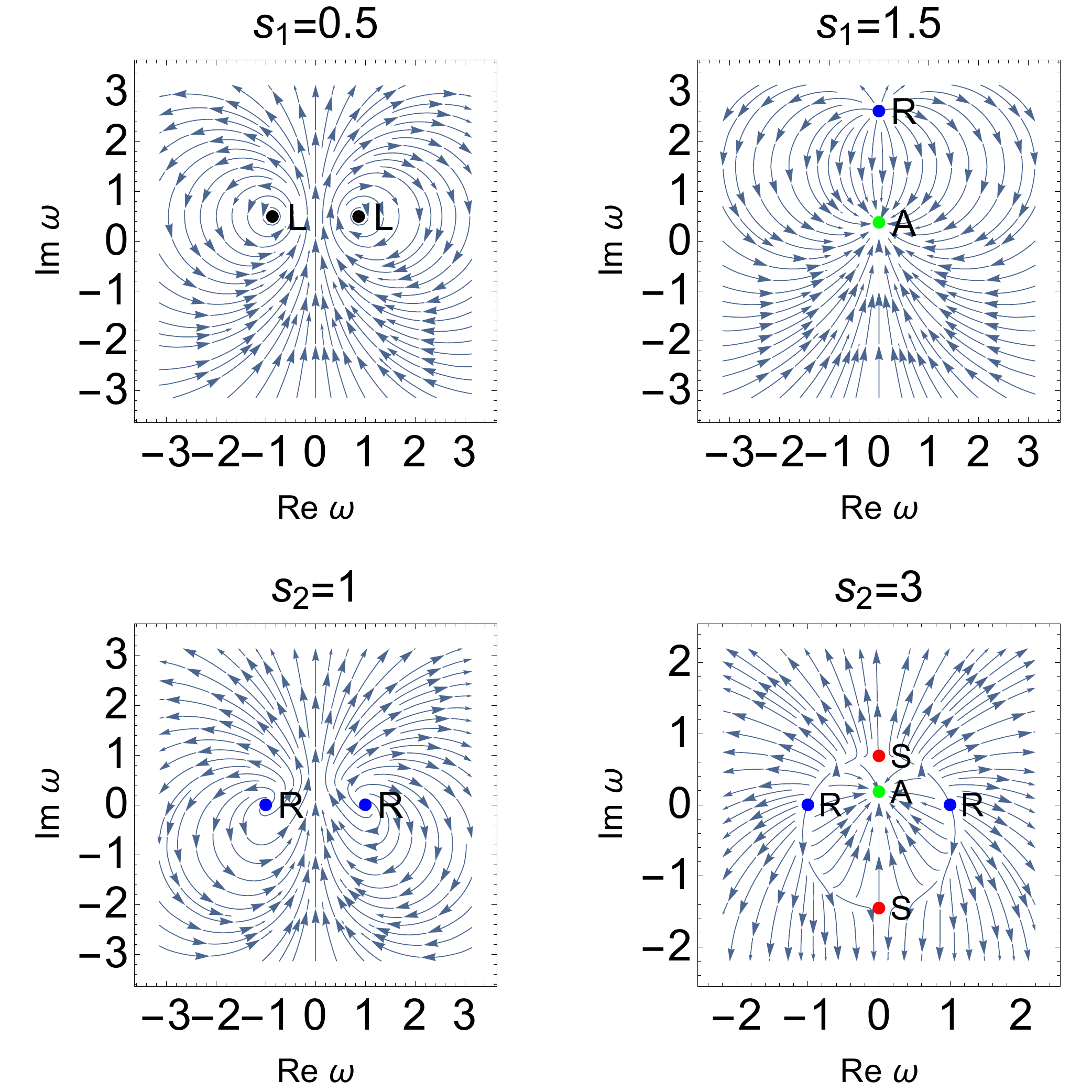}
    \caption{{\it Environmentally induced phase space transitions -- phase flows in the complex sterographic plane $w= (n_x+in_y)/(1+n_z)$}
    \\
    Bottom panel: Flow fields of Eq. \ref{eq:proj_diss} with $\gamma\to s_2$ (and equivalently, Eq. \ref{eq:var_bias} with $s_1=0$) are shown in stereographic projection . For $\gamma/J<2$ there are two repulsive fixed points of this flow at $\vec{n}= \pm \hat {\bf x}$ (marked with ``R'' and blue color) . For $\gamma/J >2$, four new fixed points emerge, two saddles (marked with ``S'' and red color) and two attractors (marked with ``A'' and green color, the bottom one being out of field of view). The great circle through the R and S fixed points splits the Bloch sphere of Fig. 1 into two mutually inaccessible hemispheres, hence the trajectory connecting the two poles is disconnected.
    \\
    Top panel: 
    Another disconnection transition, predicted from Eq. \ref{eq:var_bias} with $s_2=0$ -- see discussion in main text for the physical realization of this transition.  Formally, the similarities are due to presence/absence of the trajectory connecting the poles of the Bloch sphere on either side of the critical coupling $s_1/J=1$. There are, however differences in the details. In particular, instead of repulsive fixed points we find a family of periodic orbits and a pair of non-dynamical ``limit'' points (marked with ``L'') in the small $s_1/J<1$ regime and only a single attractive/repulsive pair of fixed points for $s_1/J>1$. }  
    \label{fig:fpoints}
\end{figure}

\noindent
{\it Mapping to pure state dynamics:}
Remarkably, Eq.~\ref{eq:lindblad} decouples completely into radial ($d$) and angular components ($|\vec{n}|=1$), with $\bar{\vec{n}} = d\vec{n}$ and corresponding density matrix evolution
\begin{align}
    \dot{\vec{n}}  &= J \hat{x} \cross \vec{n} - \gamma n_z \vec{n} \cross\left( \hat{z} \cross \vec{n}\right)
    				\nonumber\\
    				\label{eq:proj_diss}
    			&= J \hat{x} \cross \vec{n} +\gamma n_z (\hat{\vec{z}}- n_z \vec{n)}
		\\
    \dot{d} &= -\gamma(1-n_z^2),
    	\label{eq:proj_diss_radial}
	\\
    \hat\rho(t)&={\bf \hat{1}} \sqrt{1-d^2}+ d\ \vec{n} \cdot \hat{\vec{\tau}}
\end{align}
While the radial component of the dynamics quantifies the degree of thermalization of the pseudospin $\hat{\vec{\tau}}$,
angular dynamics may be thought of as describing a (non-linear) deterministic evolution of a pure state that encodes the structure of the remaining entanglement in the dimer. The spectral transition in the linear representation of the problem (Fig. \ref{fig:underover} and Eq. \ref{eq:lindblad}) manifests itself in a more dramatic transition of the fixed point structure of the angular dynamics of $\vec{n}$, whose properties and physical interpretation we focus on  in the remainder of this paper.
 
For the case considered thus far (see bottom panel of Fig. \ref{fig:fpoints}, with identification $\gamma\to s_2$) there are only two unstable (repulsive) fixed points on the underdamped side ($0<\gamma <2J$) at $\vec{n}=\pm\hat{\bf x}$. Here the trajectory of interest (see Fig.1) is the late time limiting orbit that connects the infinity ($\ket{\downarrow\uparrow}$) to the origin ($\ket{\uparrow\downarrow}$).  There are six fixed points on the overdamped side ($\gamma>2J$): in addition to the (still unstable) two at $\pm \hat{\bf x}$, there are two that are stable (attractive) and two are saddles, posessing one unstable and one stable direction each. 
The latter four new fixed points appear  
in the $yz$-plane at polar angle $\theta = 1/2 \arcsin(-2J/\gamma)$; near $2J/\gamma=1$, they emerge 
at $\phi=\pi, \theta= \pi/4 \pm (1-2J/\gamma)$ and 
$\phi=0, \theta=3\pi/4 \pm (1-2J/\gamma)$.  Importantly, separatracies passing from unstable points ($\pm x$) to saddles 
 form a circular phase boundary demarkating two mutually disconnected regions of phase space as shown in Fig. \ref{fig:fpoints}. Hence, the trajectory starting at $\ket{\uparrow \downarrow}$ never crosses this phase boundary and rather ends up in its own attractive fixed point, i.e. it becomes ``disconnected'' from $\ket{\downarrow\uparrow}$ at $\gamma \geq 2 J$. Top panel of Fig. \ref{fig:fpoints} shows another such disconnection transition 
 suggested by the statistical formalism we develop next.

{\bf Failure of Adiabatic Transport as a Dynamical Phase Transition:}

{\it Biased Trajectory Ensembles:} in classical glasses, transitions associated with dynamical properties have been analysed in terms of ensembles of trajectories. 
When extended to open~\cite{Hickey2012} and closed  quantum systems~\cite{Hickey2013}, it 
 amounts to an interpretation of the full counting
  statistics. The
  \emph{dynamical phase transition} occurs as a non-analyticity in the generating function of some time-extensive order parameter.
We now provide a general self-contained derivation of the large deviation formulation for our problem and demonstrate that the failure of this adiabatic process can be understood as such a transition.  As a byproduct of this derivation we will also be able to identify different types of such dynamical transitions.

Consider a general initial pure state $|0\rangle$ evolving under a Hamiltonian ${\hat{H}}$ and a ``biasing'' perturbation ${\hat{O}}$ of strength $s$
\begin{equation}
\partial_t \ket{\psi} 
=
\left[\hat{H} - i s \hat{O} \right]\ket{\psi}.
\label{eq:NonLinearSchrodinger}
\end{equation}  
We shall return to explain (below) how $\hat{O}$ (which may be generally state-dependent, i.e. nonlinear) is determined by post-selection of external measurements. Importantly, such evolution induces the loss of norm, with certain trajectories playing an amplified role compared to pure unitary evolution, hence the term biasing. One natural object to quantify this process is the conventional partition function associated with the time-evolved density matrix
\begin{align}
&Z_s(t)\equiv Tr |t\rangle \langle t|\\
&=\langle 0|
{\mathcal T}e^{+i \int_0^t (\hat{H}+ i \frac{s}{2} \hat{O}) dt'}
{\mathcal T}e^{-i \int_0^t (\hat{H}- i \frac{s}{2} \hat{O}) dt'} |0\rangle.
\end{align}
From here it is relatively straightforward to see that 
\begin{align}
\log Z_s(t)=-s \int_0^t \frac{\langle t'|\hat{O}|t'\rangle}{\langle t'|t'\rangle}\to -s t \langle \overline{\infty}|\hat{O}|\overline\infty\rangle,
\label{eq:logZ}
\end{align}
where the notation $|\bar{t}\rangle$ is used to denote a properly normalized counterpart to the time evolved  state $\ket{t}$, $|\bar{t}\rangle \equiv |t\rangle/\sqrt{\langle t|t\rangle}$ with corresponding (in general nonlinear) Schr\"odinger equation
$i \partial_t \ket{\bar{t}} 
=
\left[\hat{H} - is (\hat{O}-\langle\hat{O} \rangle) \right]\ket{\bar{t}}
$.
We may define the dynamic quasi-free-energy functional as
\begin{equation}
\varphi(s) = \lim_{t\rightarrow \infty} \frac{1}{t} \ln\ev{Z_z(t)}\to-s  \langle \hat{O} \rangle.\label{eq:dynamical_free_energy}
\end{equation}
Note that we tacitly assume the existence of the (unique) steady state dominating late-time averaging in both Eqs. (\ref{eq:logZ}) and (\ref{eq:dynamical_free_energy}) - more on this below.  The dynamics of observables can be obtained straightforwardly
\begin{equation}
\dot{\vec{n}} 
= i \langle  [\hat{H}_0, \hat{\vec{\tau}}]\rangle 
- \frac{s}{2} 
\langle\{ \hat{O},\hat{\vec{\tau}}\}\rangle
    + s  \langle \hat{O}\rangle \langle \hat{\vec{\tau}}\rangle .
 \label{eq:bias}
\end{equation}

\noindent
{\it Dephasing as Entanglement Bias:}

Motivated by the observed dissipation-induced suppression of entanglement above we now consider candidate biasing operators  $\hat{O}$ that may encode such an effect. Although in general the entanglement is not related to simple observables, 
here the pseudospin $\hat{\tau}_z$ is related to two standard measures of entanglement:
{i. }\ The expectation $n_z = \ev{\hat{\tau}_z} = \abs{n_1}^2-\abs{n_2}^2 = \lambda_1^2-\lambda_2^2$ is the difference between the two Schmidt coefficients for the cut across the dimer;  
{ii.}\ The variance $\ev{\delta \hat{\tau}_z^2} \equiv\ev{(\hat{\tau}_z - \ev{\tau_z})^2} = 1-n_z^2  = 2 (1-\tr \rho_A^2)=4\lambda_1^2 \lambda_2^2$ is the concurrence~\cite{Hill1997}: a lower bound to the von Neumann entanglement entropy and an entanglement monotone ($\rho_A$ is the reduced density for one spin in a dimer)

\begin{figure}[t]
a)\hspace{1.5in} b) \;\;\;\;\;\;\;\;\;\;\;\;\;\;\;\;\;\;\;\;\;\;\;\;\;\;\;
    \centering
    \includegraphics[width=0.95\linewidth, trim=0 0 0 0, clip]{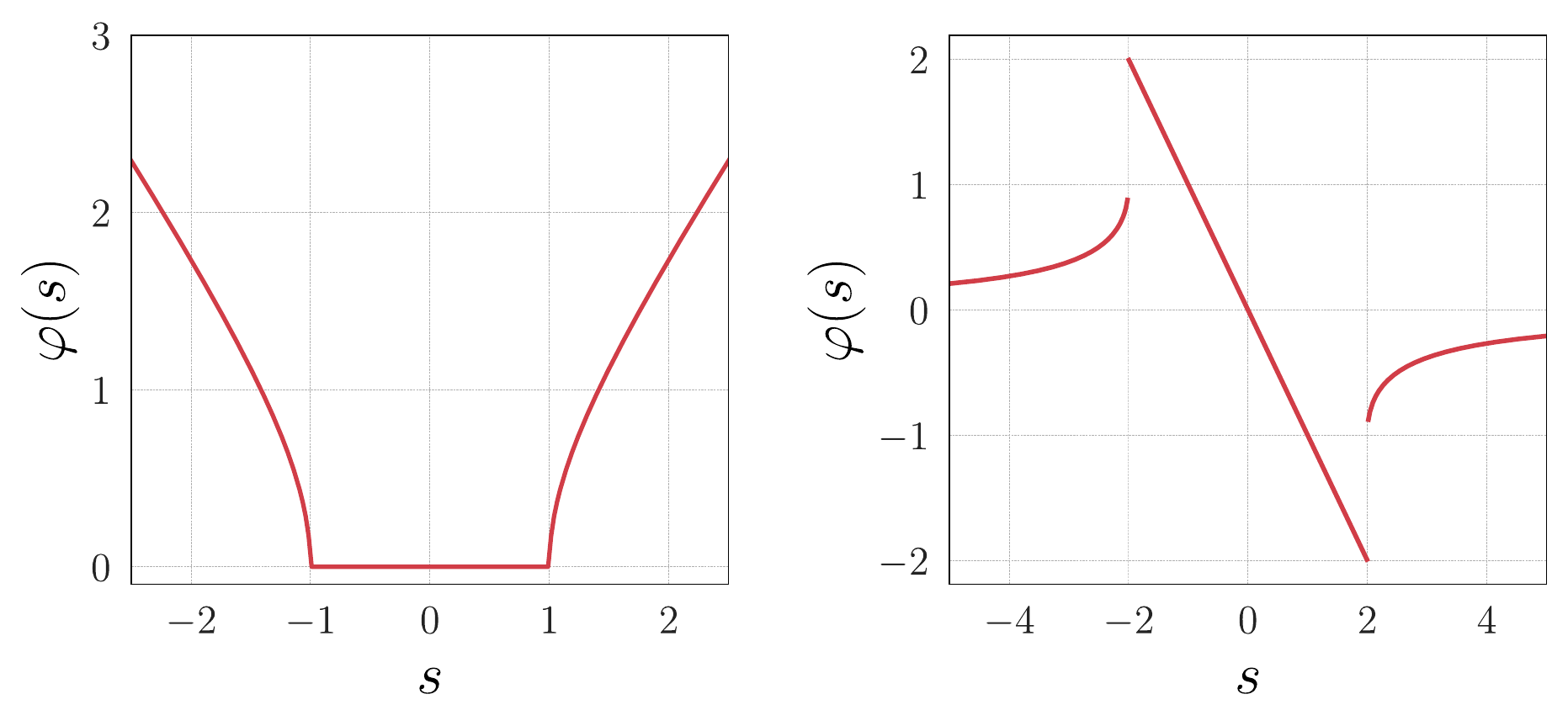}
    \caption{{\it Dynamical free energy with entanglement bias (J=1):} The dynamical free energy for the Heisenberg model 
    a) with linear bias corresponding to $s_1=s$ and $s_2=0$ in Eq.~(\ref{eq:var_bias}), and b) with variance bias corresponding to $s_1=0$ and $s_2=s$  in Eq.~(\ref{eq:var_bias}). The variance-biased cumulant generating function b) exhibits a dynamical phase transition at $\flatfrac{s}{J} = 2$, signifying the failure of the adiabatic process. The linear biased cumulant generating function a) exhibits a non-analyticity at $s/J=1$, another signature of this failure.}
    \label{fig:dynamical_phase_diagrams}
\end{figure}

We may now introduce both into Eq. \ref{eq:bias} and find
\begin{equation}
\dot{\vec{n}} 
= 
J \hat{x} \cross \vec{n} -s_1 n_z \hat{z}\cross (\hat{z} \cross \vec{n}) - s_2 n_z \vec{n} \cross (\hat{z} \cross \vec{n})
\label{eq:var_bias},
\end{equation}
which is identical to Eq.~(\ref{eq:proj_diss}) with $s_1=0$ and $s_2=\gamma$. We may also consider the other extreme case of $s_2=0$ and finite $s_1$ which is equivalent to the much studied non-Hermitian quantum mechanics problem~\cite{Hickey2013} and to postselecting trajectories for measurements of staggered magnetization. Both of these cases can be analysed along the same lines of fixed point structure (as was already done above for the variance biased case) and dynamical free energy computed, as shown in Figs. \ref{fig:fpoints} and \ref{fig:dynamical_phase_diagrams}.  
One should note that the behavior observed here is decidedely unconventional, e.g. we observe a ``zeroth order transition'' in the variance biased case.  Importantly, late time averages necessary to compute $\varphi$ are not straightforward. The underdamped regime does not possess an attractive fixed point so $n_z=0$ results from integrating over a persistent oscillation. In the overdamped regime of the variance biased case, there are two attractors (bistability). The attractor reached from the South pole of Fig. 1 (located below the field of view displayed in the bottom-right panel of Fig. \ref{fig:fpoints}) is of interest here, resulting in the discontinuous behavior of $\varphi$ in Fig. \ref{fig:dynamical_phase_diagrams}.

\vspace{0.1in}
{\bf Conclusion:}

\noindent
Dephasing biases quantum trajectories and restricts the accessible regions of Hilbert space to those with low entanglement. This can induce the failure of adiabatic quantum computation --- where intermediate states in the computation are typically highly entangled --- and of quantum control. It is mirrored in recent studies of restriction of entanglement growth in random circuits with weak or projective measurement~\cite{li2018quantum,skinner2019measurement,chan2019unitary}. We have given a concrete demonstration that --- for a simple adiabatic process in a system of two quantum spins --- this failure of adiabatic transport is a dynamical phase transition. Within a trajectory ensemble treatment, the transition is accompanied by a discontinuity in the scaled cumulant generating function associated with a time extensive order parameter. The order parameter is related to the entanglement structure of the system. 

Our model can be realised directly in coupled flux qubits. The environmental noise in such a system is often dominated by flux noise that corresponds to effective noise fields in the z-direction only~\cite{Crowley2016}. An intriguing alternative is to realise the non-linear Schr\"odinger evolution of Eq.~(\ref{eq:NonLinearSchrodinger}) using a post-selection scheme~\cite{YouPerimeter2020}. The situation that we have described effectively biases the qubit dynamics with the variance of an operator  rather than simply an operator as in the usual linear bias cases. This necessitates a greater degree of post-selection corresponding to the additional tomography required to find the variance of the operator at each time-step  or the image of this in measurements carried out on the bath~\cite{Hickey2013}. 

Can this analysis can be extended to many spins and to true adiabatic quantum computation? Quantum advantage requires as many as fifty coherent spins, and a useful paradigm for understanding computational failure must generalise to such systems.  
There are some hints that the ideas presented here can be extended successfully. Recent analysis of sweeps through a topological phase transition in the presence of an external bath reveal the same equations as Eq.~(\ref{eq:proj_diss})  albeit in a system of non-interacting particles~\cite{DecoherentQuench}. It is also possible to extend the Langevin approach used here to study more profoundly entangled many body systems~\cite{morley2019evolution} within a matrix product state Langevin description. A typical adiabatic computation undergoes several avoided crossings of low-lying levels. The states at the avoided crossings are often highly entangled. Environmental depletion of entanglement can thus prevent the avoided crossing and cause a failure of the computation. 
Thresholds for error correction in gate based models of quantum computation do not currently have an analogue in adiabatic computation. Our hope is that mapping the failure of adiabatic transport to a dynamical phase transition will prove useful in this search.

\noindent
{\it Acknowledgements:}
We thank S. Gopalakrishnan and E. Kapit for useful discussions of ideas explored in this work. VO acknowledges support from the NSF DMR Grant No. 1508538. AGG and FB acknowledge support from the EPSRC.


\begin{thebibliography}{29}
\expandafter\ifx\csname natexlab\endcsname\relax\def\natexlab#1{#1}\fi
\expandafter\ifx\csname bibnamefont\endcsname\relax
  \def\bibnamefont#1{#1}\fi
\expandafter\ifx\csname bibfnamefont\endcsname\relax
  \def\bibfnamefont#1{#1}\fi
\expandafter\ifx\csname citenamefont\endcsname\relax
  \def\citenamefont#1{#1}\fi
\expandafter\ifx\csname url\endcsname\relax
  \def\url#1{\texttt{#1}}\fi
\expandafter\ifx\csname urlprefix\endcsname\relax\def\urlprefix{URL }\fi
\providecommand{\bibinfo}[2]{#2}
\providecommand{\eprint}[2][]{\url{#2}}

\bibitem[{\citenamefont{Born and Fock}(1928)}]{born1928adiabatic}
\bibinfo{author}{\bibfnamefont{M.}~\bibnamefont{Born}} \bibnamefont{and}
  \bibinfo{author}{\bibfnamefont{V.}~\bibnamefont{Fock}},
  \bibinfo{journal}{Zeitschrift f{\"{u}}r Physik}
  \textbf{\bibinfo{volume}{51}}, \bibinfo{pages}{165} (\bibinfo{year}{1928}),
  ISSN \bibinfo{issn}{0044-3328},
  \urlprefix\url{https://doi.org/10.1007/BF01343193}.

\bibitem[{\citenamefont{Bergmann et~al.}(1998)\citenamefont{Bergmann, Theuer,
  and Shore}}]{bergmann1998k}
\bibinfo{author}{\bibfnamefont{K.}~\bibnamefont{Bergmann}},
  \bibinfo{author}{\bibfnamefont{H.}~\bibnamefont{Theuer}}, \bibnamefont{and}
  \bibinfo{author}{\bibfnamefont{B.~W.} \bibnamefont{Shore}},
  \bibinfo{journal}{Rev. Mod. Phys.} \textbf{\bibinfo{volume}{70}},
  \bibinfo{pages}{1003} (\bibinfo{year}{1998}),
  \urlprefix\url{https://link.aps.org/doi/10.1103/RevModPhys.70.1003}.

\bibitem[{\citenamefont{Kadowaki and Nishimori}(1998)}]{kadowaki1998quantum}
\bibinfo{author}{\bibfnamefont{T.}~\bibnamefont{Kadowaki}} \bibnamefont{and}
  \bibinfo{author}{\bibfnamefont{H.}~\bibnamefont{Nishimori}},
  \bibinfo{journal}{Physical Review E} \textbf{\bibinfo{volume}{58}},
  \bibinfo{pages}{5355} (\bibinfo{year}{1998}).

\bibitem[{\citenamefont{Brooke}(1999)}]{brooke1999j}
\bibinfo{author}{\bibfnamefont{J.}~\bibnamefont{Brooke}},
  \bibinfo{journal}{Science} \textbf{\bibinfo{volume}{284}},
  \bibinfo{pages}{779} (\bibinfo{year}{1999}).

\bibitem[{\citenamefont{Farhi et~al.}(2001)\citenamefont{Farhi, Goldstone,
  Gutmann, Lapan, Lundgren, and Preda}}]{farhi2001quantum}
\bibinfo{author}{\bibfnamefont{E.}~\bibnamefont{Farhi}},
  \bibinfo{author}{\bibfnamefont{J.}~\bibnamefont{Goldstone}},
  \bibinfo{author}{\bibfnamefont{S.}~\bibnamefont{Gutmann}},
  \bibinfo{author}{\bibfnamefont{J.}~\bibnamefont{Lapan}},
  \bibinfo{author}{\bibfnamefont{A.}~\bibnamefont{Lundgren}}, \bibnamefont{and}
  \bibinfo{author}{\bibfnamefont{D.}~\bibnamefont{Preda}},
  \bibinfo{journal}{Science} \textbf{\bibinfo{volume}{292}},
  \bibinfo{pages}{472} (\bibinfo{year}{2001}).

\bibitem[{\citenamefont{Santoro et~al.}(2002)\citenamefont{Santoro,
  Marto{\v{n}}{\'a}k, Tosatti, and Car}}]{santoro2002theory}
\bibinfo{author}{\bibfnamefont{G.~E.} \bibnamefont{Santoro}},
  \bibinfo{author}{\bibfnamefont{R.}~\bibnamefont{Marto{\v{n}}{\'a}k}},
  \bibinfo{author}{\bibfnamefont{E.}~\bibnamefont{Tosatti}}, \bibnamefont{and}
  \bibinfo{author}{\bibfnamefont{R.}~\bibnamefont{Car}},
  \bibinfo{journal}{Science} \textbf{\bibinfo{volume}{295}},
  \bibinfo{pages}{2427} (\bibinfo{year}{2002}).

\bibitem[{\citenamefont{Aharonov et~al.}(2008)\citenamefont{Aharonov, Van~Dam,
  Kempe, Landau, Lloyd, and Regev}}]{Aharonov2008adiabatic}
\bibinfo{author}{\bibfnamefont{D.}~\bibnamefont{Aharonov}},
  \bibinfo{author}{\bibfnamefont{W.}~\bibnamefont{Van~Dam}},
  \bibinfo{author}{\bibfnamefont{J.}~\bibnamefont{Kempe}},
  \bibinfo{author}{\bibfnamefont{Z.}~\bibnamefont{Landau}},
  \bibinfo{author}{\bibfnamefont{S.}~\bibnamefont{Lloyd}}, \bibnamefont{and}
  \bibinfo{author}{\bibfnamefont{O.}~\bibnamefont{Regev}},
  \bibinfo{journal}{SIAM review} \textbf{\bibinfo{volume}{50}},
  \bibinfo{pages}{755} (\bibinfo{year}{2008}).

\bibitem[{\citenamefont{Altshuler et~al.}(2010)\citenamefont{Altshuler, Krovi,
  and Roland}}]{Altshuler2010}
\bibinfo{author}{\bibfnamefont{B.}~\bibnamefont{Altshuler}},
  \bibinfo{author}{\bibfnamefont{H.}~\bibnamefont{Krovi}}, \bibnamefont{and}
  \bibinfo{author}{\bibfnamefont{J.}~\bibnamefont{Roland}},
  \bibinfo{journal}{Proceedings of the National Academy of Sciences}
  \textbf{\bibinfo{volume}{107}}, \bibinfo{pages}{12446}
  (\bibinfo{year}{2010}), ISSN \bibinfo{issn}{0027-8424},
  \urlprefix\url{http://www.pnas.org/content/107/28/12446}.

\bibitem[{\citenamefont{J{\"{o}}rg et~al.}(2010)\citenamefont{J{\"{o}}rg,
  Krzakala, Semerjian, and Zamponi}}]{Jorg2010}
\bibinfo{author}{\bibfnamefont{T.}~\bibnamefont{J{\"{o}}rg}},
  \bibinfo{author}{\bibfnamefont{F.}~\bibnamefont{Krzakala}},
  \bibinfo{author}{\bibfnamefont{G.}~\bibnamefont{Semerjian}},
  \bibnamefont{and} \bibinfo{author}{\bibfnamefont{F.}~\bibnamefont{Zamponi}},
  \bibinfo{journal}{Phys. Rev. Lett.} \textbf{\bibinfo{volume}{104}},
  \bibinfo{pages}{207206} (\bibinfo{year}{2010}),
  \urlprefix\url{https://link.aps.org/doi/10.1103/PhysRevLett.104.207206}.

\bibitem[{\citenamefont{Laumann et~al.}(2015)\citenamefont{Laumann, Moessner,
  Scardicchio, and Sondhi}}]{laumann2015quantum}
\bibinfo{author}{\bibfnamefont{C.~R.} \bibnamefont{Laumann}},
  \bibinfo{author}{\bibfnamefont{R.}~\bibnamefont{Moessner}},
  \bibinfo{author}{\bibfnamefont{A.}~\bibnamefont{Scardicchio}},
  \bibnamefont{and} \bibinfo{author}{\bibfnamefont{S.}~\bibnamefont{Sondhi}},
  \bibinfo{journal}{The European Physical Journal Special Topics}
  \textbf{\bibinfo{volume}{224}}, \bibinfo{pages}{75} (\bibinfo{year}{2015}).

\bibitem[{\citenamefont{Denchev et~al.}(2016)\citenamefont{Denchev, Boixo,
  Isakov, Ding, Babbush, Smelyanskiy, Martinis, and Neven}}]{Denchev2016}
\bibinfo{author}{\bibfnamefont{V.~S.} \bibnamefont{Denchev}},
  \bibinfo{author}{\bibfnamefont{S.}~\bibnamefont{Boixo}},
  \bibinfo{author}{\bibfnamefont{S.~V.} \bibnamefont{Isakov}},
  \bibinfo{author}{\bibfnamefont{N.}~\bibnamefont{Ding}},
  \bibinfo{author}{\bibfnamefont{R.}~\bibnamefont{Babbush}},
  \bibinfo{author}{\bibfnamefont{V.}~\bibnamefont{Smelyanskiy}},
  \bibinfo{author}{\bibfnamefont{J.}~\bibnamefont{Martinis}}, \bibnamefont{and}
  \bibinfo{author}{\bibfnamefont{H.}~\bibnamefont{Neven}},
  \bibinfo{journal}{Physical Review X} \textbf{\bibinfo{volume}{6}},
  \bibinfo{pages}{1} (\bibinfo{year}{2016}), ISSN \bibinfo{issn}{21603308},
  \eprint{1512.02206}.

\bibitem[{\citenamefont{Crowley et~al.}(2014)\citenamefont{Crowley, Duric,
  Vinci, Warburton, and Green}}]{Crowley2014}
\bibinfo{author}{\bibfnamefont{P.~J.~D.} \bibnamefont{Crowley}},
  \bibinfo{author}{\bibfnamefont{T.}~\bibnamefont{Duric}},
  \bibinfo{author}{\bibfnamefont{W.}~\bibnamefont{Vinci}},
  \bibinfo{author}{\bibfnamefont{P.~A.} \bibnamefont{Warburton}},
  \bibnamefont{and} \bibinfo{author}{\bibfnamefont{A.~G.} \bibnamefont{Green}},
  \bibinfo{journal}{Physical Review A} \textbf{\bibinfo{volume}{042317}},
  \bibinfo{pages}{1} (\bibinfo{year}{2014}).

\bibitem[{\citenamefont{Bauer et~al.}(2015)\citenamefont{Bauer, Wang, Pi{\v
  z}orn, and Troyer}}]{bauer2015entanglement}
\bibinfo{author}{\bibfnamefont{B.}~\bibnamefont{Bauer}},
  \bibinfo{author}{\bibfnamefont{L.}~\bibnamefont{Wang}},
  \bibinfo{author}{\bibfnamefont{I.}~\bibnamefont{Pi{\v z}orn}},
  \bibnamefont{and} \bibinfo{author}{\bibfnamefont{M.}~\bibnamefont{Troyer}}
  (\bibinfo{year}{2015}),
  \urlprefix\url{https://www.microsoft.com/en-us/research/publication/entanglement-resource-adiabatic-quantum-optimization/}.

\bibitem[{\citenamefont{Wild et~al.}(2016)\citenamefont{Wild, Gopalakrishnan,
  Knap, Yao, and Lukin}}]{wild2016adiabatic}
\bibinfo{author}{\bibfnamefont{D.~S.} \bibnamefont{Wild}},
  \bibinfo{author}{\bibfnamefont{S.}~\bibnamefont{Gopalakrishnan}},
  \bibinfo{author}{\bibfnamefont{M.}~\bibnamefont{Knap}},
  \bibinfo{author}{\bibfnamefont{N.~Y.} \bibnamefont{Yao}}, \bibnamefont{and}
  \bibinfo{author}{\bibfnamefont{M.~D.} \bibnamefont{Lukin}},
  \bibinfo{journal}{Physical review letters} \textbf{\bibinfo{volume}{117}},
  \bibinfo{pages}{150501} (\bibinfo{year}{2016}).

\bibitem[{\citenamefont{Aharonov and Ben-Or}(2008)}]{aharonov2008threshold}
\bibinfo{author}{\bibfnamefont{D.}~\bibnamefont{Aharonov}} \bibnamefont{and}
  \bibinfo{author}{\bibfnamefont{M.}~\bibnamefont{Ben-Or}},
  \bibinfo{journal}{SIAM Journal on Computing} \textbf{\bibinfo{volume}{38}},
  \bibinfo{pages}{1207} (\bibinfo{year}{2008}),
  \urlprefix\url{https://doi.org/10.1137/S0097539799359385}.

\bibitem[{\citenamefont{Jordan et~al.}(2006)\citenamefont{Jordan, Farhi, and
  Shor}}]{jordan2006error}
\bibinfo{author}{\bibfnamefont{S.~P.} \bibnamefont{Jordan}},
  \bibinfo{author}{\bibfnamefont{E.}~\bibnamefont{Farhi}}, \bibnamefont{and}
  \bibinfo{author}{\bibfnamefont{P.~W.} \bibnamefont{Shor}},
  \bibinfo{journal}{Physical Review A} \textbf{\bibinfo{volume}{74}},
  \bibinfo{pages}{052322} (\bibinfo{year}{2006}).

\bibitem[{\citenamefont{Young et~al.}(2013)\citenamefont{Young, Sarovar, and
  Blume-Kohout}}]{young2013error}
\bibinfo{author}{\bibfnamefont{K.~C.} \bibnamefont{Young}},
  \bibinfo{author}{\bibfnamefont{M.}~\bibnamefont{Sarovar}}, \bibnamefont{and}
  \bibinfo{author}{\bibfnamefont{R.}~\bibnamefont{Blume-Kohout}},
  \bibinfo{journal}{Physical Review X} \textbf{\bibinfo{volume}{3}},
  \bibinfo{pages}{041013} (\bibinfo{year}{2013}).

\bibitem[{\citenamefont{Pudenz et~al.}(2014)\citenamefont{Pudenz, Albash, and
  Lidar}}]{pudenz2014error}
\bibinfo{author}{\bibfnamefont{K.~L.} \bibnamefont{Pudenz}},
  \bibinfo{author}{\bibfnamefont{T.}~\bibnamefont{Albash}}, \bibnamefont{and}
  \bibinfo{author}{\bibfnamefont{D.~A.} \bibnamefont{Lidar}},
  \bibinfo{journal}{Nature communications} \textbf{\bibinfo{volume}{5}},
  \bibinfo{pages}{1} (\bibinfo{year}{2014}).

\bibitem[{\citenamefont{Crowley and Green}(2016)}]{Crowley2016}
\bibinfo{author}{\bibfnamefont{P.~J.~D.} \bibnamefont{Crowley}}
  \bibnamefont{and} \bibinfo{author}{\bibfnamefont{A.~G.} \bibnamefont{Green}},
  \bibinfo{journal}{Physical Review A} \textbf{\bibinfo{volume}{94}},
  \bibinfo{pages}{1} (\bibinfo{year}{2016}), ISSN \bibinfo{issn}{24699934}.

\bibitem[{\citenamefont{Kamenev}(2011)}]{kamenev2011field}
\bibinfo{author}{\bibfnamefont{A.}~\bibnamefont{Kamenev}},
  \emph{\bibinfo{title}{Field theory of non-equilibrium systems}}
  (\bibinfo{publisher}{Cambridge University Press}, \bibinfo{year}{2011}).

\bibitem[{\citenamefont{Hickey et~al.}(2012)\citenamefont{Hickey, Genway,
  Lesanovsky, and Garrahan}}]{Hickey2012}
\bibinfo{author}{\bibfnamefont{J.~M.} \bibnamefont{Hickey}},
  \bibinfo{author}{\bibfnamefont{S.}~\bibnamefont{Genway}},
  \bibinfo{author}{\bibfnamefont{I.}~\bibnamefont{Lesanovsky}},
  \bibnamefont{and} \bibinfo{author}{\bibfnamefont{J.~P.}
  \bibnamefont{Garrahan}}, \bibinfo{journal}{Physical Review A - Atomic,
  Molecular, and Optical Physics} \textbf{\bibinfo{volume}{063824}},
  \bibinfo{pages}{1} (\bibinfo{year}{2012}).

\bibitem[{\citenamefont{Hickey et~al.}(2013)\citenamefont{Hickey, Genway,
  Lesanovsky, and Garrahan}}]{Hickey2013}
\bibinfo{author}{\bibfnamefont{J.~M.} \bibnamefont{Hickey}},
  \bibinfo{author}{\bibfnamefont{S.}~\bibnamefont{Genway}},
  \bibinfo{author}{\bibfnamefont{I.}~\bibnamefont{Lesanovsky}},
  \bibnamefont{and} \bibinfo{author}{\bibfnamefont{J.~P.}
  \bibnamefont{Garrahan}}, \bibinfo{journal}{Physical Review B - Condensed
  Matter and Materials Physics} \textbf{\bibinfo{volume}{87}},
  \bibinfo{pages}{1} (\bibinfo{year}{2013}), ISSN \bibinfo{issn}{10980121},
  \eprint{1211.4773}.

\bibitem[{\citenamefont{Hill and Wootters}(1997)}]{Hill1997}
\bibinfo{author}{\bibfnamefont{S.}~\bibnamefont{Hill}} \bibnamefont{and}
  \bibinfo{author}{\bibfnamefont{W.~K.} \bibnamefont{Wootters}},
  \bibinfo{journal}{Physical Review Letters} \textbf{\bibinfo{volume}{78}},
  \bibinfo{pages}{5022} (\bibinfo{year}{1997}), ISSN \bibinfo{issn}{10797114},
  \eprint{9703041}.

\bibitem[{\citenamefont{Li et~al.}(2018)\citenamefont{Li, Chen, and
  Fisher}}]{li2018quantum}
\bibinfo{author}{\bibfnamefont{Y.}~\bibnamefont{Li}},
  \bibinfo{author}{\bibfnamefont{X.}~\bibnamefont{Chen}}, \bibnamefont{and}
  \bibinfo{author}{\bibfnamefont{M.~P.} \bibnamefont{Fisher}},
  \bibinfo{journal}{Physical Review B} \textbf{\bibinfo{volume}{98}},
  \bibinfo{pages}{205136} (\bibinfo{year}{2018}).

\bibitem[{\citenamefont{Skinner et~al.}(2019)\citenamefont{Skinner, Ruhman, and
  Nahum}}]{skinner2019measurement}
\bibinfo{author}{\bibfnamefont{B.}~\bibnamefont{Skinner}},
  \bibinfo{author}{\bibfnamefont{J.}~\bibnamefont{Ruhman}}, \bibnamefont{and}
  \bibinfo{author}{\bibfnamefont{A.}~\bibnamefont{Nahum}},
  \bibinfo{journal}{Physical Review X} \textbf{\bibinfo{volume}{9}},
  \bibinfo{pages}{031009} (\bibinfo{year}{2019}).

\bibitem[{\citenamefont{Chan et~al.}(2019)\citenamefont{Chan, Nandkishore,
  Pretko, and Smith}}]{chan2019unitary}
\bibinfo{author}{\bibfnamefont{A.}~\bibnamefont{Chan}},
  \bibinfo{author}{\bibfnamefont{R.~M.} \bibnamefont{Nandkishore}},
  \bibinfo{author}{\bibfnamefont{M.}~\bibnamefont{Pretko}}, \bibnamefont{and}
  \bibinfo{author}{\bibfnamefont{G.}~\bibnamefont{Smith}},
  \bibinfo{journal}{Physical Review B} \textbf{\bibinfo{volume}{99}},
  \bibinfo{pages}{224307} (\bibinfo{year}{2019}).

\bibitem[{\citenamefont{You et~al.}(2020{\natexlab{a}})\citenamefont{You, Kuo,
  Arovas, and Vishveshwara}}]{YouPerimeter2020}
\bibinfo{author}{\bibfnamefont{Y.-Z.} \bibnamefont{You}},
  \bibinfo{author}{\bibfnamefont{W.-T.} \bibnamefont{Kuo}},
  \bibinfo{author}{\bibfnamefont{D.}~\bibnamefont{Arovas}}, \bibnamefont{and}
  \bibinfo{author}{\bibfnamefont{S.}~\bibnamefont{Vishveshwara}},
  \bibinfo{journal}{Perimeter Quantum Frontiers Seminar}
  (\bibinfo{year}{2020}{\natexlab{a}}).

\bibitem[{\citenamefont{You et~al.}(2020{\natexlab{b}})\citenamefont{You, Kuo,
  Arovas, and Vishveshwara}}]{DecoherentQuench}
\bibinfo{author}{\bibfnamefont{Y.-Z.} \bibnamefont{You}},
  \bibinfo{author}{\bibfnamefont{W.-T.} \bibnamefont{Kuo}},
  \bibinfo{author}{\bibfnamefont{D.}~\bibnamefont{Arovas}}, \bibnamefont{and}
  \bibinfo{author}{\bibfnamefont{S.}~\bibnamefont{Vishveshwara}},
  \emph{\bibinfo{title}{Decoherent quench acros quantum phase transitions}},
  \bibinfo{howpublished}{http://pirsa.org/20110004}
  (\bibinfo{year}{2020}{\natexlab{b}}).

\bibitem[{\citenamefont{Morley-Wilkinson}(2019)}]{morley2019evolution}
\bibinfo{author}{\bibfnamefont{J.~G.} \bibnamefont{Morley-Wilkinson}}, Ph.D.
  thesis, \bibinfo{school}{UCL (University College London)}
  (\bibinfo{year}{2019}).

\end{thebibliography}
\end{document}